\newcommand{\be}{\begin{equation}}\newcommand{\ee}{\end{equation}}
\newcommand{\bea}{\begin{eqnarray}}\newcommand{\eea}{\end{eqnarray}}
\newcommand{\nn}{\nonumber}
\newcommand{\lb}[1]{\label{#1}}

\def\tr{{\rm tr}}

\newcommand\cN{{\cal N}}

\def\nb{\nabla}
\def\sB{\stackrel{\frown}{\Box}}

\def\f{\frac}

\documentclass[12pt]{article}
\usepackage{amsmath,amssymb}
\usepackage{cite}

\numberwithin{equation}{section}

\begin{document}
\begin{titlepage}
\begin{flushright}
\end{flushright}
\vspace{0.7cm}

\begin{center}
{\Large\bf Supersymmetry at BLTP: Recent Progress \\
\vspace{0.1cm}

in Two Directions}

\vspace{1.5cm}

{\large\bf I.L. Buchbinder and E.A. Ivanov} \\
\vspace{1cm}

 {\it Bogoliubov  Laboratory of Theoretical Physics, JINR,
141980 Dubna, Russia} \\
{\tt buchbinder@theor.jinr.ru, eivanov@theor.jinr.ru }

\end{center}
\vspace{1cm}

\begin{abstract}
\noindent Ten years ago, in a paper \cite{60}, a brief historical survey
of the research activity in the Sector ``Supersymmetry'' at the Bogoliubov Laboratory of Theoretical Physics (BLTP)
for more than 50 years of its
existence has been given. Here, in commemoration of the 70th jubilee of Joint Institute for Nuclear Research, we review some recent sound advancements in this area. Specifically, we consider the issues of
constructing the superfield quantum effective actions in $6D, {\cal N}=(1,0)$ supersymmetric gauge theories and off-shell unconstrained superfield formulations of ${\cal N}=2$ higher spins.
In both cases, the harmonic superspace approach
plays the decisive role.
\end{abstract}

\end{titlepage}
\tableofcontents
\newpage

\section{Introduction}
The  paper \cite{60} was devoted to the brief description of the mainstream scientific activity of the Sector N2 ``Supersymmetry''  of BLTP for more than fifty years passed since its foundation.
It was stressed there that the most influential pioneering results and methods which have successfully passed the examination by time were the following: {1)} notoph
as the first example of antisymmetric gauge field \cite{notoph}; {2)}``Ogievetsky Theorem'' \cite{Teorem} and the view of the gravitation theory as a theory of spontaneous breaking,
with the graviton as a Goldstone field \cite{BorOg}; {3)} the inverse Higgs phenomenon in nonlinear
realizations \cite{InvH}; {4)} the complex superfield geometry of ${\cal N}=1$ supergravity \cite{ESsg,ESsg1};
{5)} the general relationship between linear and nonlinear realizations of supersymmetry
\cite{IKa2,IKa3}; {6)} Grassmann analyticity and harmonic superspace \cite{GAn,HSS,HSS1,4,4-1,book}. As the future prospective directions of research, in \cite{60} there
were mentioned the exploring of the geometry and quantum
structure of supersymmetric gauge theories and supergravity in diverse dimensions in the superfield approach and the study of various aspects of supersymmetric and superconformal mechanics models in their intertwining
relationships with the higher-dimensional field theories and string theory.

Here we will concentrate on the following two themes which have received a considerable attention for the last 10 years: the entirely new domain
of using the harmonic superspace approach for off-shell description of
the higher-spin $4D, {\cal N}=2$ supersymmetric theories, as well as the further applications of the quantum harmonic ${\cal N}=2$ formalism for calculations
of the effective action of $6D$ supersymmetric gauge theories. We would like to notice in advance that in this short review of our activity for the last decade we as a rule refer
to some basic papers on the subject, while the more detailed corpus of references can be found in our original works
\footnote{There were many other interesting and sound findings in the supersymmetry-related areas at BLTP for the last decade (see, e.g., \cite{SFed,KBTV}).}.

\section{${\cal N}=2$ Higher Spins from Harmonic Approach}
In this Section we briefly review our approach to constructing the manifest $4D,\,\mathcal{N}=2$ supersymmetric higher spin field theories.
The theories are formulated in $\mathcal{N}=2$ harmonic superspace that provides manifest off-shell supersymmetry combined with the explicit gauge invariance.
Analytic superfields carrying $\mathcal{N}=2$ massless supermultiplets of higher spins are described and their interactions with the hypermultiplet are presented.
The models constructed were further generalized to involve the superconformal couplings. We also discuss how to define $\mathcal{N}=2$ higher
spin models on AdS background within the harmonic formalism.

The plan of the Section is as follows. In subsection 2.1 we recall basic notions of $4D, {\cal N}=2$ harmonic superspace. In subsection 2.2 and 2.3 the structure of analytic gauge potentials
describing off-shell $4D, {\cal N}=2$ higher-spin multiplets is explained and their couplings to $q^+$ hypermultiplets are presented. Subsection 2.4 is devoted to generalization to the superconformal case.
In subsections 2.5 and 2.6 the AdS$_4$ version of harmonic superspace is defined and the simplest examples of models invariant under $4D, {\cal N}=2$ AdS supersymmetry are given.
Subsection 2.7 contains summary and outlook.

\subsection{$4D$ Harmonic Superspace: why does it matter?}
Supersymmetry, despite lacking its experimental confirmation so far, is in the heart of the modern mathematical and quantum physics. It allowed to construct a lot of new theories with remarkable features: supergravities, superstrings, superbranes, ${\cal N}=4$ super
  Yang-Mills (the first example of
  the UV-finite quantum field theory), etc. It also exhibited unexpected relations between them, e.g.,
  {``gravity/gauge''} duality.

The natural approach to supersymmetric theories is the superfield methods. The natural generalization of Minkowski space {$x^m$} to supersymmetry is {${\cal N}$} {\bf extended Minkowski superspace}
\bea {\cal M}^{(4|4{\cal N})} = \left( x^m\,,\;\theta^\alpha_i\,, \;\bar\theta^{\dot\alpha\,i}\right), \; i =1, \ldots, {\cal N}\,,\nonumber \eea
where {$\theta^\alpha_i\,,\bar\theta^{\dot\alpha\,i}$} are Grassmann coordinates,
{$\{\theta, \theta\}, \{\theta, \bar\theta\}= 0$}.

Supersymmetric theories are adequately formulated off shell in terms of superfields defined on various superspaces.
The basic {$4D, {\cal N}=1$} superspace is chiral superspace, with half the original {${\cal N}=1$} Grassmann coordinates. Its {${\cal N}=2$} analog
is {\bf Harmonic Superspace}. Indeed, in  four dimensions, the only self-consistent off-shell superfield formalism for both {${\cal N}=2$} and {${\cal N}=3$} theories
is known to be the harmonic superspace approach \cite{HSS1,4,book}.

Harmonic {${\cal N}=2$} superspace is defined as the coordinate set
\bea
Z = (x^m\,,\;\theta^\alpha_i\,, \;\bar\theta^{\dot\alpha\,j}, u^{\pm i}), \quad u^{\pm i} \in SU(2)/U(1),  \; u^{+ i}u^-_i = 1\,.
   \lb{HSS}
\eea
The basic merit of such an extension is the existence of the analytic harmonic {${\cal N}=2$} superspace involving half the number of Grassmann coordinates:
\bea
\zeta_A = (x^m_A, \theta^{+ \alpha}, \bar\theta^{+
\dot\alpha}, u^{\pm i}), \; \theta^{+ \alpha, \dot\alpha} := \theta^{\alpha, \dot\alpha i} u^+_i, \;
x^m_A := x^m - 2i\theta^{(i}\sigma^m \bar\theta^{j)}u^+_iu^-_j\,.
   \lb{AnalS}
\eea
All basic {${\cal N}=2$} superfields are analytic:
\bea
\underline{\rm SYM}:&& V^{++}(\zeta_A); \;\;  \underline{\rm matter \;hypermultiplets}: \; q^{+}(\zeta_A)\,, \,\bar{q}^{+}(\zeta_A); \nonumber \\
\underline{\rm supergravity}:&& H^{++ m}(\zeta_A)\,,\,H^{++ \hat\alpha +}(\zeta_A)\,, \,H^{++ 5}(\zeta_A)\,, \hat\alpha = (\alpha, \dot\alpha)\,.  \nonumber
\eea

An instructive example is Abelian {${\cal N}=2$} gauge theory,
\bea
V^{++}(\zeta_A)\,, \quad \delta V^{++} = D^{++}\Lambda (\zeta_A)\,, \; D^{++} = \partial^{++} - 4i\theta^{+\alpha}\bar\theta^{+\dot\alpha}\partial_{\alpha\dot\alpha}. \nonumber
\eea
In Wess-Zumino gauge, the analytic gauge prepotential  $V^{++}$ involves just 8 + 8 off-shell degrees of freedom which constitute ${\cal N}=2$ vector multiplet:
\bea
&& V^{++}(\zeta_A) = (\theta^+)^2 \phi + (\bar\theta^+)^2 \bar\phi + 2i\theta^{+\alpha}\bar\theta^{+\dot\alpha} A_{\alpha\dot\alpha} \nonumber \\
&& +\, (\bar\theta^+)^2 \theta^{+\alpha}\psi_\alpha^i u^-_i +
(\theta^+)^2 \bar\theta^{+}_{\dot\alpha}\bar\psi^{\dot\alpha i} u^-_i +(\theta^+)^2(\bar\theta^+)^2 D^{(ik)}u^-_iu^-_k\,. \nonumber
\eea
The invariant action is written with making use of the non-analytic gauge connection $V^{--}$:
\bea
&& S \sim \int d^{12}Z\,\, V^{++} V^{--}\,, \; D^{++} V^{--} - D^{--} V^{++} = 0\,, \; \delta V^{--} = D^{--}\Lambda\,, \nonumber \\
&& [D^{++}, D^{--}] = D^0, \quad D^0 V^{\pm\pm} = \pm 2\,V^{\pm\pm}\,. \nonumber
\eea
\subsection{Supersymmetry and higher spins}
 Supersymmetric higher-spin theories provide
 a bridge between superstring theory and low-energy (super)gauge theories.

 The free massless bosonic and fermionic higher spin field theories were constructed in
 \cite{Fr,FrFang}. The component approach to $4D$, ${\cal N}=1$ supersymmetric
free massless higher spin models was worked out in \cite{Courtright1979,Vasiliev1980}.
The complete off-shell {${\cal N}=1$} superfield Lagrangian formulation of {$4D, {\cal N}=1$}
free higher spins was given in \cite{Kuz93,Kuz94}.
 At the same time, an off-shell superfield Lagrangian formulation for higher-spin {\bf extended}
 supersymmetric theories, with all supersymmetries manifest,  was unknown for long
even for free theories.

This gap was filled in \cite{IBIZ1}.
An off-shell manifestly {${\cal N}=2$} supersymmetric unconstrained formulation of {$4D, {\cal N}=2$} super Fronsdal theory for integer spins was constructed  in the harmonic superspace approach.
The general case with the maximal integer spin ${\bf s}$ is spanned by the analytic gauge potentials
\bea
h^{++\alpha(s-1)\dot\alpha(s-1)}(\zeta), h^{++\alpha(s-2)\dot\alpha(s-2)}(\zeta), h^{++\alpha(s-1)\dot\alpha(s-2)+}(\zeta),
h^{++\dot\alpha(s-1)\alpha(s-2)+}(\zeta), \lb{GenMult}
\eea
where {$\alpha(s) := (\alpha_1 \ldots \alpha_s), \dot\alpha(s) := (\dot\alpha_1 \ldots \dot\alpha_s)$}
The relevant gauge transformations can also be defined and shown to leave, in the WZ-like gauge, the  physical field multiplet
{$({\bf s, s-1/2, s-1/2, s-1})$} plus some auxiliary fields.
 The on-shell spin contents of {${\cal N}=2$} higher-spin multiplets can be summarized as\footnote{The superspin of the given supermultiplet
 is defined as the largest spin of the component fields.}
\bea
&&\underline{superspin\; 1}: \; 1, \,(1/2)^2,\, (0)^2\,, \nonumber \\
&&\underline{superspin\; 2}: \; 2,\, (3/2)^2,\, 1 \,,\nonumber  \\
&&\underline{superspin \;3}: \; 3, \,(5/2)^2, \, 2\,, \nonumber \\
&& ....... \nonumber \\
&&\underline{superspin \;s}: \; s, \, (s-1/2)^2, \, s-1\,. \nonumber
\eea

The manifestly {${\cal N}=2$} supersymmetric off-shell cubic couplings of the integer superspin {$4D, {\cal N}=2$} gauge multiplets to the matter hypermultiplets were
further constructed in \cite{IBIZ2,IBIZ2a}.
Quite recently, the harmonic superspace non-conformal construction was generalized
to the case of {${\cal N}=2$} superconformal multiplets and their hypermultiplet couplings \cite{IBIZ4}, as well as
to the AdS$_4$ case \cite{IvaZai}\footnote{A possible way to extend the whole construction to half-integer superspins was proposed in \cite{IvZai2}.}.

\subsection{Hypermultiplet couplings}

The construction of interactions in the theory of
higher spins is a very important (albeit difficult) task (see, e.g.,\cite{MMRu2011}).

In \cite{IBIZ2} we have constructed the off-shell
manifestly {$\mathcal{N}=2$} supersymmetric cubic couplings
$(\mathbf{\frac{1}{2}, \frac{1}{2}, s})$  of an arbitrary higher
integer  superspin {$\mathbf{s}$} gauge {$\mathcal{N}=2$} multiplet to the
hypermultiplet matter in {$4D, \mathcal{N}=2$} harmonic
superspace. In our approach {$\mathcal{N}=2$} supersymmetry of cubic vertices is always manifest and off-shell, in contrast,
e.g., to the non-manifest light-cone formulations \cite{Mets1,Mets2}.

The starting point is the {${\cal N}=2$} hypermultiplet off-shell free action:
\begin{equation}
    S = \int  d\zeta^{(-4)}  \; \mathcal{L}^{+4}_{free} = -\int d\zeta^{(-4)}  \; \frac{1}{2} q^{+a} \mathcal{D}^{++} q^+_a, a = 1,2\,. \lb{qAct}
\end{equation}
The analytic gauge potentials for any spin ${\bf s}$ with the correct transformation rules are recovered by proper gauge-covariantization
of the harmonic derivative {$\mathcal{D}^{++}$}. The simplest option is gauging of {$U(1)$},
\begin{eqnarray}
&&\delta q^{+a} = -\lambda_0 J q^{+ a}, \quad J q^{+ a} = i (\tau_3)^a_{\;b} q^{+b}\,,\nonumber \\
&& {\cal D}^{++} \;\Rightarrow \;{\cal D}^{++} + \hat{\cal H}^{++}_{(1)}\,, \quad  \hat{\cal H}^{++}_{(1)}= h^{++}J\,, \nonumber \\
&& \delta_\lambda \hat{\cal H}^{++}_{(1)} = [{\cal D}^{++}, \hat{\Lambda}]\,, \quad \hat{\Lambda}= \lambda J\; \Rightarrow \;
\delta_\lambda h^{++} = {\cal D}^{++}\lambda \,.\nonumber
\eea
In {${\cal N}=2$} supergravity, that is for {${\bf s}=2$},

\begin{eqnarray}
&&    S_{(2)} =  -\int d\zeta^{(-4)}  \; \frac{1}{2} q^{+a} \big(\mathcal{D}^{++} + {\cal H}_{(2)}\big)q^+_a, \quad \delta{\cal H}_{(2)}
 = [\mathcal{D}^{++}, \hat{\Lambda}_{(2)}],\lb{SGact} \\
&& \quad {\cal H}_{(2)}= h^{++ M}(\zeta)\partial_M, \; \hat{\Lambda}_{(2)}
= \lambda^{M}(\zeta)\partial_M, \; M := (\alpha\dot\beta, 5, \hat{\mu}+)\,. \nonumber
\end{eqnarray}

For higher ${\bf s}$ everything goes analogously. E.g., for {${\bf s}=3$}
\begin{eqnarray}
&&    S_{(3)} =  -\int d\zeta^{(-4)}  \; \frac{1}{2} q^{+a} \big(\mathcal{D}^{++} + {\cal H}_{(3)}J \big)q^+_a, \lb{spin3Act}\\
&&\delta{\cal H}_{(3)} = [\mathcal{D}^{++}, \hat{\Lambda}_{(3)}], \quad {\cal H}_{(3)}
= h^{++\alpha\dot\alpha\,M}(\zeta)\partial_M\partial_{\alpha\dot\alpha}, \quad \hat{\Lambda}_{(3)}
= \lambda^{\alpha\dot\alpha\,M}(\zeta)\partial_M\partial_{\alpha\dot\alpha}\,. \nonumber
\end{eqnarray}

\subsection{Superconformal couplings}

Free conformal higher-spin actions in {$4D$} Minkowski space were
pioneered in \cite{FraTseyt,FraLin1,FraLin2}. Since then,
a lot of works on (super)conformal higher spins followed (e.g., \cite{Segal2003,KuzenkoManv2017}, and many others).
(Super)conformal higher-spin theories are considered as a basis for all other types of higher-spin models. Non-conformal ones follow from the superconformal ones through couplings
to the {\bf superfield compensators}.

In \cite{IBIZ4}, the  off-shell $4D, \mathcal{N}=2$ higher spins and their hypermultiplet couplings were extended to the superconformal case, once again based
on the harmonic Grassmann  analyticity principle.
Rigid {$4D, \mathcal{N}=2$} superconformal symmetry plays a crucial role in fixing the structure of the theory.

{$4D, \mathcal{N}=2$} superconformal algebra (SCA) preserves harmonic analyticity and is a closure of the rigid {$\mathcal{N}=2$} supersymmetry and special conformal symmetry
\bea && \delta_\epsilon \theta^{+ \hat{\alpha}} = \epsilon^{\hat{\alpha}i}u^+_i\,, \; \delta_\epsilon x^{\alpha\dot\alpha} =
-4i \left( \epsilon^{\alpha i} \bar{\theta}^{+\dot{\alpha}} + \theta^{+\alpha} \bar{\epsilon}^{\dot{\alpha}i} \right) u^-_i\,, \hat{\alpha} = (\alpha, \dot\alpha) \,,\nonumber \\
&&\delta_k \theta^{+ {\alpha}} = x^{\alpha\dot\beta}k_{\beta\dot\beta}\theta^{\hat{\beta}}\,, \;\delta_k x^{\alpha\dot\alpha}
= x^{\rho\dot\alpha}k_{\rho\dot\rho}x^{\dot\rho\alpha}\,, \; \delta_k u^{+ i} =
(4i\theta^{+\alpha}\bar\theta^{+\dot\alpha}k_{\alpha\dot\alpha}) u^{- i}\,. \nonumber
\eea

What about the conformal properties of various analytic higher-spin potentials? No problems with the spin ${\bf 1}$ potential {$V^{++}$}:
\bea \delta_{sc} V^{++} = - \hat\Lambda_{sc}V^{++}\,, \quad \hat\Lambda_{sc}:= \lambda^{\alpha\dot\alpha}_{sc}\partial_{\alpha\dot\alpha}
+ \lambda^{\hat\alpha+}_{sc}\partial_{\hat\alpha+} +
\lambda^{++}_{sc}\partial^{--}\,. \lb{Vconf} \eea
The cubic vertex {$\sim q^{+a}V^{++}Jq^+_a$} is invariant up to total derivative provided that

\bea \delta_{sc} q^{+a}  = -\hat\Lambda_{sc} q^{+a} - \frac12 \Omega q^{+a}\,, \quad \Omega :=
    (-1)^{P(M)} \partial_M \lambda^M\,. \lb{qConf}
\eea

The  situation gets more complicated for ${\bf s}\geq 2$. Requiring {${\cal N}=2$} gauge potentials for ${\bf s}=2$ to be closed under
{${\cal N}=2$} SCA necessarily leads to
\bea
&& {\cal D}^{++} \rightarrow {\cal D}^{++} + \kappa_2 \hat{\mathcal{H}}^{++}_{(s=2)}\,, \nonumber \\
&& \hat{\mathcal{H}}^{++}_{(s=2)} : = h^{++M} \partial_M
    =
        h^{++\alpha\dot{\alpha}}\partial_{\alpha\dot{\alpha}}
        +
    h^{++\alpha+}\partial^-_\alpha
    +
    h^{++\dot{\alpha}+} \partial^-_{\dot{\alpha}}
    +
    h^{(+4)}\partial^{--} \nonumber \\
&&     \delta_{k_{\alpha\dot{\alpha}}} h^{(+4)} = - \hat{\Lambda} h^{(+4)}
    +
    4i h^{++\alpha+} \bar{\theta}^{+\dot{\alpha}} k_{\alpha\dot{\alpha}}
    +
    4i \theta^{+\alpha} h^{++\dot{\alpha}+}
    k_{\alpha\dot{\alpha}}\nonumber
\eea
For ensuring conformal covariance, it is impossible to avoid introducing the extra potential {$h^{(+4)}$}. This extended
set of potentials embodies {${\cal N}=2$} {\bf Weyl multiplet}
({${\cal N}=2$} conformal SG gauge multiplet).

 For {${\bf s}\geq 3$} the gauge-covariantization of the free {$q^{+ a}$} action requires adding the gauge superfields carried by differential operators
of rank ${\bf s -1}$ in {$\partial_M$},
\bea
{\cal D}^{++} \rightarrow {\cal D}^{++} + \kappa_s \hat{\mathcal{H}}^{++}_{(s)}(J)^{P(s)}\,, \quad P(s) = \frac{1 + (-1)^{s-1}}{2}\,. \nonumber
\eea
In particular, for ${\bf s}=3$:
\bea \hat{\cal H}_{(s=3)}= h^{++MN}\partial_N\partial_M + h^{++}, \quad  h^{++MN} = (-1)^{P(M)P(N)} h^{++ NM}\,.\lb{s3H}
\eea
The whole consideration can be repeated for the general integer higher-spin {$s$} case. The relevant generalized Weyl multiplets involve
 $ 8(2s -1)_B + 8(2s-1)_F$ d.o.f. off shell.

 The self-consistent ${\cal N}=2$ superconformal couplings of higher-spin gauge superfields
 to the matter $q^+$ hypermultiplets at the full nonlinear level and for an arbitrary ${\cal N}=2$ conformal supergravity background
 were also presented for the first time in \cite{IBIZ4}. It is worth noting that the higher-spin ${\cal N}=2$ gauge supermultiplets  in the description through
 the Mezincescu-type prepotentials for the particular case of conformally flat backgrounds were earlier considered in \cite{KuRa}.

\subsection{Towards AdS background}

In view of the celebrated {$AdS/CFT$} correspondence, it is of high importance to explicitly construct {${\cal N}=2$} higher spins in the AdS background,
with the superconformal symmetry {$SU(2,2|2)$} being broken
to the AdS supersymmetry {$OSp(2|4;R)$}.

One way to achieve this is to start from the covariant formalism in the {$AdS_4, {\cal N}=2$} superspace defined as the coset {$OSp(2|4;R)/[SO(2) \times SL(2, C)]$},
thus generalizing the {$AdS_4, {\cal N}=1$} superfield approach of \cite{IvSor}.  This way was chosen  in ref.\cite{KuTMa2008}, essentially based on the so called
projective superspace techniques and treating the case of $5D, {\cal N}=1$ AdS superspace (see also a recent paper \cite{BuKuSa}).

The approach of ref. \cite{IvaZai} is distinguished in that it exclusively proceeds from the analyticity-preserving realization of the superconformal symmetry in {${\cal N}=2$} harmonic superspace
and identifies {${\cal N}=2$} AdS supersymmetry {$OSp(2|4;R)$} as its subalgebra, {$OSp(2|4;R) \subset SU(2,2|2)$}. So the super AdS supersymmetry
is already implied by the superconformal
symmetry. Once again, the manifest ${\cal N}=2$ harmonic Grassmann analyticity plays the decisive role.

 The embedding of {${\cal N}=2$} AdS superalgebra into {$SU(2,2|2)$} is realized through the identification \cite{1001,BILS202}
\bea
&&\Psi^i_\alpha = Q^i_\alpha + c^{ik} S_{k\alpha}, \quad \bar{\Psi}^i_{\dot\alpha} = \overline{\Psi^i_\alpha} = \bar{Q}_{\dot\alpha i} + c_{ik}\bar{S}^k_{\dot\alpha}, \lb{Embed}\\
&& c^{ik} = c^{ki}\, \qquad \overline{c^{ik}} = c_{ik} = \varepsilon_{il}\varepsilon_{kj} c^{lj}\,.
\eea
The {$SU(2,2|2)$} commutation relations imply for super AdS generators

\bea
&& \{ \Psi^i_\alpha,  \Psi^k_\beta \} = -4c^{ik} L_{(\alpha\beta)} +4 i \varepsilon_{\alpha\beta}\varepsilon^{ik} I, \quad I := c_{lm}T^{lm}\,,
\quad [I, \Psi^i_\alpha] = c^{ik}\Psi_{k\alpha}\,, \nonumber \\
&& \{ \Psi^i_\alpha,  \bar\Psi_{\dot\beta k}\} = 4\delta^i_k\,R_{\alpha\dot\beta}\,, \;\; R_{\alpha\dot\beta} = P_{\alpha\dot\beta} + \frac12 c^2\, K_{\alpha\dot\beta}\,, \;\;
c^2 := c^{ik}c_{ik} \sim \frac{1}{R^2_{AdS}}, \nonumber \\
&& [R_{\alpha\dot\alpha}, R_{\gamma\dot\gamma}] = \frac12 c^2\big(\varepsilon_{\alpha\gamma} \bar{L}_{\dot\alpha\dot\gamma} + \varepsilon_{\dot\alpha \dot\gamma} L_{\alpha\gamma} \big),
\quad [R_{\alpha\dot\beta},\Psi^i_\beta]  = \varepsilon_{\alpha\beta}\,\bar\Psi^i_{\dot\beta}\;\, ({\rm and \,c.c.})\,. \lb{AdSalg}
\eea

The \textbf{super AdS} transformation can be then read off as :

\begin{equation}
    \begin{split}
& \delta_\epsilon x^{\alpha\dot\alpha} = -4i\, \big[\epsilon^{\alpha i}\bar\theta^{+\dot\alpha} + \theta^{+\alpha}\bar\epsilon^{\dot\alpha i}
- c^{ik} \big(x^{\alpha\dot\beta}\bar\epsilon_{\dot\beta k} \bar\theta^{+\dot\alpha} + x^{\beta\dot\alpha}\epsilon_{\beta k}\theta^{+\alpha}\big)\big] u^-_i , \\
& \delta_\epsilon \theta^{+ \alpha}= \big(\epsilon^{\alpha i} - x^{\alpha\dot\alpha} c^{ik}\bar\epsilon_{\dot\alpha k}\big) u^+_i - 2i(\theta^+)^2 c^{ki}\epsilon^\alpha_k u^-_i\,,  \\
& \delta_\epsilon \bar\theta^{+ \dot\alpha}= \big(\bar\epsilon^{\dot\alpha i}+ x^{\alpha\dot\alpha} c^{ik}\epsilon_{\alpha k} \big) u^+_i
+ 2i(\bar\theta^+)^2 c^{ik}\bar\epsilon^{\dot\alpha}_k u^-_i,  \\
&
\delta_\epsilon u^{+i} = -4i \big[ c^{kl} u^+_k  (\epsilon_{\alpha l}\theta^{+ \alpha} + \bar\epsilon_{\dot\alpha l}\bar\theta^{+\dot\alpha})\big] u^{-i} \,.
\end{split}\lb{SuperAdS}
\end{equation}

The \textbf{nonlinear AdS} translations look as:

\begin{equation}
    \begin{split}
& \delta_{a} x^{\alpha\dot\alpha} = a^{\alpha\dot\alpha} + \frac12 c^2 a_{\beta\dot\beta}x^{\alpha\dot\beta} x^{\beta\dot\alpha}
= a^{\alpha\dot\alpha} \big(1 - \frac14 c^2 x^2\big) + \frac12 c^2(ax) x^{\alpha\dot\alpha}\,, \\
& \delta_{a} \theta^{+\alpha} = \frac12 c^2 a_{\beta\dot\beta} x^{\alpha\dot\beta} \theta^{+\beta}
=
\frac{1}{4} c^2 (ax) \theta^{+\alpha}
+
\frac{1}{2} c^2 x^{(\alpha \dot\beta} a_{\beta)\dot\beta} \theta^{+\beta} \,, \\
&
 \delta_{a} \bar\theta^{+\dot\alpha}
= \frac12 c^2 a_{\beta\dot\beta} x^{\beta\dot\alpha} \bar\theta^{+\dot\beta}
=
\frac{1}{4} c^2 (ax) \bar{\theta}^{+\dot{\alpha}}
+
\frac{1}{2} c^2 x^{\beta(\dot\alpha} a_{\beta\dot{\beta})} \bar{\theta}^{+\dot{\beta}}
\, , \\
& \delta_{a} u^{+i} = 2i  \left(c^2 a_{\alpha\dot\alpha} \theta^{+\alpha}\bar\theta^{+ \dot\alpha} \right) u^{-i}\,.
\end{split} \lb{NonlTran}
\end{equation}

The AdS supersymmetry transformation of the analytic integration measure reads:
\bea
\delta_{\epsilon} d\zeta^{(-4)} = 8i \left( c^{kl}u^-_k \epsilon_{\alpha l} \theta^{+\alpha}\right) d\zeta^{(-4)}\,. \nonumber
\eea
To compensate this variation of the integration measure in the action of {$q^{+a}$}, the latter should include the appropriate weight factor in its transformation,
\be
\delta_{\epsilon} q^{+ a} = - 4i \left(c^{kl}u^-_k\epsilon_{\alpha l} \theta^{+\alpha} \right) q^{+a}. \lb{AdSq}
\ee
The harmonic derivative $\mathcal{D}^{++}$ transforms under AdS supersymmetry as
\be
\delta_\epsilon D^{++} = 4i\,c^{kl}u^+_k\epsilon_{\alpha l} \theta^{+\alpha}\,D^0\,, \quad D^0 q^{+a} = q^{+a}, \; q^{+a}q^{+}_{a} = 0\,.\nonumber
\ee
Then the free {$q^+$} action,
\bea
S_{(2)} =  -\int d\zeta^{(-4)}  \; \frac{1}{2} q^{+a} \mathcal{D}^{++}q^+_a\,, \nonumber
\eea
is invariant under  the {${\cal N}=2$} superconformal transformations and hence under the {${\cal N}=2$} super AdS ones.
The question is how to construct the actions invariant only under {$OSp(2|4;R)$} symmetry. To accomplish this, one needs to properly break
{$SU(2,2|2)$}.

\subsection{Two ways of constructing {$OSp(2|4;R)$} invariants}

We start with the {${\cal N}=2$} AdS invariant mass term. The basic idea is to extend
the coordinate action of generators of AdS supersymmetry
by extra ``matrix'' pieces of an external {$SO(2)$} realized on  {$q^{+a}$}.
 In the flat limit such a modified {${\cal N}=2$} AdS  contracts into the central-charge extended
{$4D, {\cal N}=2$} supersymmetry.

Denoting formally this extra {$SO(2)$} generator as {$\partial_5$} and introducing
\be
q^{+} (\zeta, x^5) = e^{-im_q x^5} q^{+}(\zeta)\,, \quad \tilde{q}^{+} (\zeta, x^5) = e^{im_q x^5} \tilde{q}^{+}(\zeta)\,, \quad [m_q] = l^{-1}\,,
\ee
find the realization of {$OSp(2|4;R)$} by the following Killing super-vector
\bea
&&    \hat{\Lambda}_{AdS} = \lambda_{AdS}^{\alpha\dot{\alpha}}\partial_{\alpha\dot{\alpha}} + \lambda^{\hat{\alpha}+}_{AdS} \partial^-_{\hat{\alpha}}
    +
    \lambda^{++}_{AdS} \partial^{--}
    +
    \lambda^5_{AdS} \partial_5, \quad \lambda^5_{AdS} =\lambda^5 _\epsilon + \lambda^5_{so(2)}\,, \nonumber \\
&&\lambda^5 _\epsilon  = 2i e(x) \Big[ (\theta^+ \epsilon^-) - x^{\alpha\dot{\alpha}}\bar{\theta}^+_{\dot{\alpha}}( \epsilon^+_\alpha c^{--} - \epsilon^-_\alpha y)\Big]\times \nonumber \\
&&    \times \Big\{1 - 2i \left[(\theta^+)^2 - (\bar{\theta}^+)^2 \right] e(x)  c^{--}  \Big\} + (c.c.), \nonumber \\
&&\lambda^5_{so(2)} =
 \frac{\gamma}{2}
 +
  i \left[(\theta^+)^2 -  (\bar{\theta}^+)^2\right]   \gamma c^{--} e(x)
 -
 6  (\theta^+)^4\gamma   \left( c^{--} \right)^2  e(x)^2, \lb{sKill}
\eea
where {$y = c^{ij} u^+_iu^-_j, e(x) = \frac{1}{1+ m^2x^2/2}$}.

In order to construct an invariant {$q^+$} action, one needs to lengthen {${\cal D}^{++}$} as
\bea
&& {\cal D}^{++} \;\Rightarrow \; {\cal D}^{++} + H^{++5}_{AdS} \partial_5\,, \nonumber \\
&& H^{++5}_{AdS}  =i \left[(\theta^+)^2 -  (\bar{\theta}^+)^2\right] e(x)
    -
    6(\theta^+)^4 e(x)^2 \,c^{--}, \; \delta_{AdS} H^{++5}_{AdS} = {\cal D}^{++}\lambda^5_{AdS}\,. \nonumber
\eea

After elimination of an infinite tail of the auxiliary fields and integrating over Grassmann and harmonic variables,
the bosonic part of the massive AdS hypermultiplet action can be written as
\begin{equation}
    \begin{split}
        S_{scalar} &= - \frac{1}{2} \int d\zeta^{(-4)} \; q^{+a} \left( \mathcal{D}^{++} + H_{AdS}^{++5} \partial_5 \right) q^+_a\; \vert_{bos}
        \\&=
        \int d^4x \; \Big(  \partial_n f^i \partial^n \bar{f}_i  - m_q^2 e(x)^2 f^i \bar{f}_i
        -
     im_q e(x)^2 f^i \bar{f}^j c_{(ij)}
        \Big),
    \end{split} \lb{ScalAct}
\end{equation}
where we observe the presence of the explicit {$SU(2)$} breaking part in the mass term, with the residual {$SO(2)$} invariance.

 After rescaling {$f^i \to e(x) \hat{f}^i$} we obtain the AdS covariant scalar field
\begin{equation}
    \delta_a \hat{f}^i (x) = 0
\end{equation}
and gain the standard kinetic action for the massive scalars given on AdS metric $g_{mn} = e(x)^2 \eta_{mn}$
(with the scalar curvature  {$R = 4\Lambda =-48 c^{ik}c_{ik}$}):
\begin{equation}
    S_{scalar} = \int d^4x \sqrt{-g} \left( g^{mn} \partial_m \hat{f}^i \partial_n\bar{\hat{f}}_i
    +
    \frac{R}{6} \hat{f}^i \bar{\hat{f}}_i
    - m^2_q \hat{f}^i \bar{\hat{f}}_i
        -
    im_q  \hat{f}^i \bar{\hat{f}}^j c_{(ij)} \right). \lb{ScalAct2}
\end{equation}

 After passing to the special frame {$c_{ij} = m \delta_{ij}$}, diagonalizing mass matrix and rewriting {$\hat{f}^i = (f_+, f_-)$},
one obtains the splitting of mass for the scalar fields proportional to the parameter {$m$},

\begin{equation}
    m_\pm^2 = m_q^2 \pm m m_q\,. \nonumber
\end{equation}

For the fermionic fields one needs a similar redefinition, {$\psi_\alpha(x) \to e(x)^{\frac{3}{2}} \hat{\psi}_\alpha(x)$}. Using
the AdS covariant derivative {$\nabla_{\alpha\dot\alpha} = \big(1 + \frac{m^2}{2} x^2\big)\partial_{\alpha\dot\alpha}
= e(x)^{-1} \partial_{\alpha\dot{\alpha}}$}, one comes to the action for Dirac spinor on AdS space:
\bea
&&    S_{fer} = \int d^4x \sqrt{-g} \Big\{
     i \bar{\hat{\psi}}^{\dot{\alpha}} \left(\nabla_{\dot\alpha}^\alpha - \frac{3}{2} m^2 x^\alpha_{\dot{\alpha}} \right) \hat{\psi}_\alpha +
    i \bar{\hat{\kappa}}^{\dot{\alpha}} \left(\nabla_{\dot\alpha}^\alpha - \frac{3}{2} m^2 x^\alpha_{\dot{\alpha}} \right)  \hat{\kappa}_\alpha \nonumber \\
&&  \quad\quad\quad  +\, \frac{m_q}{2}\big(\hat{\psi}^\alpha \hat{\kappa}_\alpha
    + \bar{\hat{\psi}}_{\dot\alpha} \bar{\hat{\kappa}}^{\dot\alpha}\big) \Big\}.
\eea
So, {$m_f = m_q, \quad m_+^2 + m_-^2 = 2m^2_f$}, like in {${\cal N}=1$} AdS case \cite{IvSor}.

In order to produce a more general class of ${\cal N}=2$ AdS invariant systems one needs to redefine the superfield {$q^{+a}$} in such a way that it transforms
as a scalar superfield with zero weight under the AdS subgroup of {${\cal N}=2$} superconformal group. This is the superfield Weyl-type transformation  preserving {${\cal N}=2$}
harmonic analyticity.

We start from the free {$q^{+}$} action, {$S_{free} = -\frac12\int d\zeta^{(-4)}q^{+a}\mathcal{D}^{++} q^+_a$}. It
is superconformally invariant and hence is invariant under super {${\cal N}=2$} super AdS$_4$ group. Then one makes a Weyl-type rescaling of {$q^{+}$},

\bea
q^{+a} = G^{\frac12}\,\hat{q}^{+a}\,, \; G = \frac{\Big(1 + \frac{y^2}{m^2}\Big)}{\Big(1 + \frac{m^2 x^2}{2}\Big)^2} \Big(1 + \theta\,terms\Big), \; y:= c^{+-} = c^{ik}u^+_iu^-_k\,,\lb{Weyl}
\eea
so that $\hat{q}^{+a}$ is a scalar of zero weight under ${\cal N}=2$ super AdS$_4$ group. The {$\hat{q}^{+}$} action takes the form manifestly invariant under this group

\bea
S_{free} = -\frac12\int d\zeta^{(-4)}\,G \,\hat{q}^{+a}\mathcal{D}^{++} \hat{q}^+_a\,, \quad \delta_{AdS}\hat{q}^{+a} = 0\,. \nonumber
\eea

The new integration measure  {$d \zeta^{(-4)} G$} is invariant  under {$OSp(2|4;R)$}. So one can add to the Lagrangian  any proper function of {$\hat{q}^{+a}$}
without breaking of {$OSp(2|4;R)$}. In particular, one can add an arbitrary {${\cal L}^{+4}(\hat{q}^{+a}, u^-)$} and so gain a wide class
of the hyper-K\"ahler type sigma-model actions
on the AdS$_4$ background.

To be more precise, one can consider the following action

\begin{equation}
    S^{AdS}_{HK} = \int d\zeta^{(-4)} \Sigma \, \left[ \hat{q}^+_a \mathcal{D}^{++} \hat{q}^{+a} + L^{(+4)} (\hat{q}^+, w^+, u^-) \right],\lb{GenActAdS}
\end{equation}
where
\be
w^{+i} := u^{+ i} - u^{- i} c^{++} \frac{y}{y^2 + m^2}\,, \qquad \delta_{AdS}w^{+i} = 0\,,\nonumber
\ee
is a new harmonic variable which is inert under {$OSp(2|4)$}. In the properly defined flat limit this action goes over just to the general
action of hyper-K\"ahler {${\cal N}=2$} sigma models \cite{HSS}. It generically does not exhibit any external isometry (besides the intrinsic {$SO(2)$}),
quite similar to its flat {${\cal N}=2$} counterparts. It would be interesting to find out the relevant deformation of the hyper-K\"ahler geometry
and to clarify the role of the breaking parameter {$c^{ik}$} in it.

\subsection{Summary and outlook}
The theory of {${\cal N}= 2$} supersymmetric higher spins  {$s\geq 3$} opened a new promising direction
of applications of the harmonic superspace approach which earlier proved to be indispensable for description of
more conventional {${\cal N}= 2$} theories with maximal spins {$s\leq 2$}. Once again,
the basic property underlying these new higher-spin theories {\bf is the harmonic Grassmann analyticity} (all basic gauge potentials
are unconstrained analytic superfields involving an infinite number of degrees of freedom off shell before fixing WZ-type gauges).
As we saw on the example of hypermultiplet, harmonic superspace admits a nice generalization to the case of AdS background where
{\bf the harmonic Grassmann analyticity principle} again plays the decisive role. We expect that the same principle will efficiently work
as well for various {${\cal N}= 2$} AdS supergravities and their higher-spin analogs, and this will allow us to get further insights into the geometric and
quantum structure of these notable theories. One of the most interesting problems here is to construct AdS analogs of analytic prepotenials
of diverse {${\cal N}=2$} supergravities on the flat background \footnote{The explicit expressions for the linearized conserved higher-spin {${\cal N}=2$}
supercurrents were presented in a recent paper \cite{NZsolo}.} and to find the higher-spin counterparts of these AdS analytic prepotentials.

\section{Effective Actions of $6D$ Supergauge Theories in Harmonic Superspace}
In this Section we review the basic results of our study of the quantum structure of six-dimensional rigid supersymmetric gauge theories. We consider the harmonic superspace formulation of the six-dimensional $\cN=(1,0)$ and
$\cN=(1,1)$ rigid supersymmetric field theories and present the manifestly $\cN=(1,0)$ supersymmetric and gauge-invariant methods of constructing
off-shell quantum effective action for these theories. It is shown that in
$\cN=(1,1)$ theory, the one-loop effective action is finite off shell and the off-shell divergences of the two-loop effective action are proportional to the classical equations of motion.
Also the results regarding the finite parts of the one-loop low-energy effective action and the divergent structure of the higher-derivative $\cN=(1,0)$ supergauge theories are briefly outlined.

\subsection{Motivations}
Over past few years, our group in BLTP has carried out a series of
studies of the quantum structure of supersymmetric gauge theories in
six dimensions
\cite{BIMS-1,BIMS-2,BIMS-3,BIM-1,BIMS-4,BIMS-5,BIMS-6,BIMS-7,BIM-2,BIMS-8,BIMS-9,BIMS-10,BBIMS,BIMS-11}.
In this Section  of our survey  we briefly describe the methods and approaches developed in our works and
the results obtained by making use of these techniques.

The modern interest in the higher-dimensional supersymmetric field
theories is inspired by superstring theory. The specific feature of
the latter is the existence of the so called $D$-branes which amount to the $D+1$ dimensional surfaces in the ten-dimensional
space-time. In the low-energy limit the $D$ brane is associated with
$D+1$-dimensional supersymmetric gauge theory (see e.g., \cite{CJ}).
Therefore, the low-energy limit of superstring theory can be
described by (extended) supersymmetric quantum field theory in diverse
dimensions.

Another motivation to study the higher-dimensional extended
supergauge theories is associated with M-theory (see, e.g., \cite{BBSh}).
The hypothetical  $M$-theory is characterized by two extended objects:
$M2$-brane and $M5$-brane in eleven-dimensional space-time. The field
description of interacting multiple $M2$-branes is given by
Bagger-Lambert-Gustavsson theory which is $3D$, $\cN=8$
supersymmetric gauge theory (see the review \cite{BL}). It is
expected that the interacting multiple $M5$-branes are related to some
new \textbf{six}-dimensional extended supersymmetric gauge theory.

The study of the quantum structure of six-dimensional supersymmetric
gauge theories dimensionally reduced from superstrings was initiated
in works \cite{SW}, \cite{S}. The one- and two-loop divergences in
these theories were studied in refs.
\cite{FT,HS,MS,MS1,HS1,Kaz,Bork} \footnote{Like in the previous Sections, we basically give here
only those references which are directly related to our studies. The
full list of references can be found in
\cite{BIMS-1,BIMS-2,BIMS-3,BIM-1,BIMS-4,BIMS-5,BIMS-6,BIMS-7,BIM-2,BIMS-8,BIMS-9,BIMS-10,BBIMS,BIMS-11}.}.
We wish to pay attention to the fact that practically all
considerations in these works have been carried out in the on-shell
component approach. The basic goal of our works
\cite{BIMS-1,BIMS-2,BIMS-3,BIM-1,BIMS-4,BIMS-5,BIMS-6,BIMS-7,BIM-2,BIMS-8,BIMS-9,BIMS-10,BBIMS,BIMS-11}
was to develop the approach based on off-shell superfields and to
study the off-shell structure of quantum effective action in $6D$
supersymmetric gauge theories.

For description of the six-dimensional supergauge theories we use
the harmonic superfield method. This approach was initially
worked out for formulating $4D$ extended supersymmetric theories in
terms of unconstrained $\cN=2$ superfields \cite{HSS1} (see \cite{book} for details
and applications). Harmonic superfield approach was
generalized to six dimensions  in refs. \cite{HSW,Z,BIS,ISZ}. It
allows us to construct the supersymmetric models of gauge multiplet coupled to hypermultiplet
in terms of unconstrained $\cN=(1,0)$
harmonic superfields \cite{BIS}. If the hypermultiplet belongs to
adjoint representation, such a model possesses an additional hidden
on-shell $\cN=(0,1)$ supersymmetry and actually describes the maximally
extended rigid six-dimensional supersymmetric gauge theory.

The Section is organized as follows. In subsection 3.2 we briefly describe
$6D,\, \cN=(1,0)$ harmonic superspace and the corresponding $6D,\,
\cN=(1,0)$ and $6D,\, \cN=(1,1)$ supergauge theories in terms of
$\cN=(1,0)$ harmonic superfields. In subsection 3.3 we discuss the
background field method in harmonic superspace which allows one to
construct the manifest $\cN=(1,0)$ supersymmetric and gauge
invariant quantum effective action. In subsection 3.4 we describe
calculations of the one-loop divergences in general $\cN=(1,0)$ model
and show that in $\cN=(1,1)$ supersymmetric theory the one-loop divergences vanishes
off-shell. In subsection 3.5 we describe calculations of the two-loop
off-shell divergences and show that they are proportional to
classical equations of motion. Subsection 3.6 is devoted to deriving the one-loop low-energy effective
action in $\cN=(1,1)$ supersymmetric theory. In subsection 3.7 we develop a method to calculate the
one-loop divergences in higher-derivative $\cN=(1,0)$ supergauge theory. In the concluding subsection 3.8 we summarize
the results obtained.

\subsection{$6D, \cN=(1,1)$ supersymmetric gauge theory}
Six-dimensional superalgebra is characterized by
two independent supercharges forming non-equivalent $6D$ spinors (see e.g., \cite{HST}). The simplest
subsets of this algebra are denoted as $\cN=(1,0)$ and
$\cN=(0,1)$ that allows to formulate the $\cN=(1,1)$ supergauge
theory as the theory invariant both under $\cN=(1,0)$ and under $\cN=(0,1)$
supersymmetries. Such a theory is a maximally extended rigid
supergauge theory in six dimensions.

The harmonic superspace description of six-dimensional supersymmetry has
been given in \cite{HSW,Z,BIS}. In this approach,
$\cN=(1,1)$ theory is formulated as a model of analytic gauge
superfield $V^{++}$ coupled to hypermultiplet $q^{+}_A$ in adjoint
representation of gauge group (see the details
in \cite{BIS}). The classical action of such a theory is manifestly
$\cN=(1,0)$ supersymmetric by construction and has the form
 \bea
 S_0[V^{++}, q^+] &=& \frac{1}{\rm
f_0^2}\Big\{\sum\limits^{\infty}_{n=2} \frac{(-i)^{n}}{n} \tr \int
d^{14}z\, du_1\ldots du_n \frac{V^{++}(z,u_1 ) \ldots
V^{++}(z,u_n ) }{(u^+_1 u^+_2)\ldots (u^+_n u^+_1 )}  \nn \\
&& -\f12  \tr \int d\zeta^{(-4)} \, {q}^{+A}
\nabla^{++}q^{+}_A\Big\}, \label{S0}
 \eea
where $\nb^{++}=D^{++} + i V^{++}$  and  ${\rm f_0}$ is a dimensional coupling constant ($[{\rm f_0}]=m^{-1}$). The analytic superfield $V^{++}$, describing the gauge multiplet takes the values in the gauge group algebra,
 \bea
V^{++}=(V^{++})^A T^A,  \qquad [T^A, T^B] = i f^{ABC} T^C, \qquad
A,B,C=1,..,d_G\,, \label{Vfirst}
 \eea
where $f^{ABC}$ are totally antisymmetric structure constants
and $d_G$ is dimension of the gauge group.  The generators $T^A_{\rm
F}=t^A$ are normalized in the standard way, $\tr( t^At^B ) = \tfrac12
\delta^{AB}$.

The action (\ref{S0}) is invariant under the following gauge
transformations
 \bea
 (V^{++})' = e^{i\lambda^A T^A} V^{++}
 e^{-i\lambda^A T^A} -i e^{i\lambda^A T^A}D^{++} e^{-i\lambda^A T^A},
 \qquad ( q^{+})' =  e^{i\lambda^A T^A }q^{+},
 \label{gtr}
 \eea
where gauge parameter $\lambda^A(\zeta, u)$ is a real analytic
superfield.

Let us introduce the real non-analytic gauge connection $V^{--}=
(V^{--})^A T^A$ satisfying the harmonic zero-curvature equation \cite{book}
\bea
 D^{++} V^{--} - D^{--}V^{++} + i[V^{++},V^{--}]=0\,.
 \label{zeroc}
 \eea
Using $V^{--}$ one can define the analytic superfield
$F^{++}$ \cite{ISZ,BIS} as
 \begin{equation}
 \label{identity00}
F^{++} \equiv (D^+)^4 V^{--}, \qquad \nb^{++} F^{++}=0\,.
 \end{equation}

The classical equations of motion for the model with action (\ref{S0}) read
 \bea
E^{++}= F^{++} + \f{i}{2}\, [q^{+A}, q^{+}_A] = 0\,,  \qquad
\nabla^{++} q^{+} = 0\,. \label{eqm}
  \eea
The action (\ref{S0}) possesses an additional hidden on-shell
$\cN=(0,1)$ supersymmetry
\begin{equation}
\label{Hidden}
\delta_{(0,1)} V^{++} = \epsilon^{+ A}q^+_A\,, \quad \delta_{(0,1)}
q^{+}_A = -(D^+)^4 (\epsilon^-_A V^{--})\,, \quad \epsilon^{\pm}_A =
\epsilon_{aA}\theta^{\pm a}\,.
\end{equation}
The supersymmetry (\ref{Hidden}) mixes the  gauge multiplet and
hypermultiplet \cite{BIS}, when hypermultiplet belongs to the same
adjoint representation as the gauge multiplet.

\subsection{Background field method}
Background field method is a procedure to construct quantum
effective action in gauge theories in the form preserving classical
gauge invariance in quantum theory. For $6D,\, \cN=(1,0)$ supergauge
theories such a method was developed in our works
\cite{BIMS-1,BIMS-2,BIMS-3}.

In the background field method we split the superfield
$V^{++}$ into the sum of the ``background'' superfields $V^{++},\,
Q^{+}$ and the ``quantum'' ones $v^{++},\,q^{+}$,
    \bea
V^{++}\to {V}^{++} + {\rm f}_0 v^{++}; \qquad q^{+} \to Q^{+} + q^{+}.
    \eea
Then we expand the effective action in a power series in quantum
superfields and obtain a theory of the superfields $v^{++},\ q^{+}$
in the background of the classical superfields $V^{++},\, Q^{+}$,
which are treated as functional arguments of the effective action.
Our aim is to study the two-loop contributions to the effective
action in the gauge superfield sector. To this end, it is sufficient
to assume that the hypermultiplet is purely quantum.

Using the results of refs. \cite{BIMS-1,BIMS-2,BIMS-3}, the general
expression for the effective action can be written in the form
\bea
\label{Effective_Action} e^{i \Gamma[V^{++},Q^{+}]} =
\mbox{Det}^{1/2}\sB \int {\cal D}v^{++}\,{\cal D}q^+\, {\cal D}
b\,{\cal D} c\,{\cal D}\varphi\, \exp\Big(iS_{total}\Big),
\eea
where the operator $\sB=\frac{1}{2}(D^+)^4(\nb^{--})^2$, when acting on a
space of analytic superfields, is reduced to the covariant superfield
d'Alembertian
\begin{eqnarray}
\label{Box_First_Part} \sB = \eta^{MN} \nabla_M \nabla_N + i W^{+a}
\nabla^{-}_a + i F^{++} \nabla^{--} - \frac{i}{2}(\nabla^{--}
F^{++}),
\end{eqnarray}
and $\eta_{MN}$ is $6D$ Minkowski metric with the mostly negative
signature. The total action $S_{tot}$ has the form
\bea S_{tot} =
\tilde{S} + S_{gf} + S_{FP} + S_{NK}\,,
\eea
and it includes the
gauge-fixing term corresponding to the Feynman gauge,
\bea
S_{gf}[v^{++}, V^{++}] = -\frac{1}{2}\tr \int d^{14}z du_1
du_2\,\frac{v_\tau^{++}(1)v_\tau^{++}(2)}{(u^+_1u^+_2)^2} \\+
\frac{1}{4}\tr \int d^{14}z du\, v_\tau^{++} (D^{--})^2
v_\tau^{++}\,, \label{SGF}
\eea
the action $S_{FP}$ for the
fermionic Faddeev-Popov ghosts $b$ and $c$, as well as the action
$S_{NK}$ for the bosonic real analytic Nielsen-Kallosh ghost
$\varphi$,
\bea S_{FP} &=&-\tr\int d\zeta^{(-4)} \, \nb^{++} b\,
(\nb^{++} c +i[v^{++}, c]), \label{FP}\\
S_{\mbox{\scriptsize NK}} &=& -\frac{1}{2}\tr \int d\zeta^{(-4)} \,
\varphi ({\nb}^{++})^2\varphi \label{NK}.
\eea
The label $\tau$ of $v_\tau^{++}$ means that the
superfield is taken in $\tau$-frame \cite{book}. The action
(\ref{SGF}) depends on the background field $V^{++}$ through the
background gauge bridge superfield. The action $\tilde{S} = S[V^{++}+{\rm f}_{0}v^{++},Q^{+}+q^{+}]-
S[V^{++},Q^{+}] - \int\, S'_{V^{++}}{\rm f}_{0}v^{++}-\int\, S'_{Q^{+}}q^{+}.$ Here $S'_{V^{++}}$ and $S'_{Q^{+}}$ are the
functional derivative of the action with respect to $V^{++}$ and $Q^{+}$, respectively
\footnote{The expressions $S'_{V^{++}}{\rm f}_{0}v^{++}$ and $S'_{Q^{+}}q^{+}$ are integrated over harmonic superspace.}.  This means that
the expansion of $\tilde{S}$ with respect of quantum superfields starts with quadratic terms.

\subsection{One-loop divergences}
The calculation of the effective action is carried out in the
framework of the loop expansion. In the one-loop approximation the
quantum corrections to the classical action are determined by the
quadratic part of the action $S_{tot}$. After integration over
quantum superfields this quadratic part produces the one-loop
contribution $\Gamma^{(1)}$ to the effective action. In the case under consideration
after some transformations one gets the following expression for one-loop contribution to the effective
action
\bea
\label{one-loop}
\Gamma^{(1)}[V^{++},Q^{+}]= \frac{i}{2}Tr\ln(\sB-2{\rm f}_{0}^2 Q^{+}G_{(1,1)}Q^{+})
- \frac{i}{2}Tr\ln(\sB)\\
\nonumber - iTr\ln(\nabla^{++})_{adj}^2 +
\frac{i}{2}Tr\ln(\nabla^{++})_{adj}^2 +iTr\ln(\nabla^{++})_{R}^2.
\eea
Here $Tr$ is a functional trace in superspace, $G_{(1,1)}$ is Green function for the operator
$\nabla^{++}$ (see all details in \cite{BIMS-2}) and the subscripts  $adj$
and $R$ mean the adjoint representation and an arbitrary $R$ representations for the
hypermultiplets.

Calculating the one-loop divergences of superfield functional
determinants is accomplished in the framework of the proper-time
technique (a superfield version of Schwinger-De Witt technique). Such
a technique allows us to preserve the manifest gauge invariance and
manifest $\cN=(1,0)$ supersymmetry at all steps of calculations.

The general scheme of calculations can schematically be summarized as
\begin{itemize}
\item{Proper-time representation}
$$
Tr\ln O \sim Tr \int_0^\infty
\frac{d(is)}{(is)^{1+\varepsilon}}e^{isO_{1}}\delta(1,2)|_{2=1}\,.
$$
\item{Here $s$ is the proper-time parameter and $\varepsilon$ is a
parameter of dimensional regularization.}
\item{Generically, the total delta-function $\delta(1,2)$ contains
the Grassmann delta-function $\delta^8(\theta_1 - \theta_2)$, which vanishes at
$\theta_1=\theta_2$}.
\item{Typically, the operator $O$ contains some number of spinor derivatives $D^{+}_{a}, D^{-}_{a}$ which act
on the Grassmann delta-functions $\delta^8(\theta_1 - \theta_2)$ and
can kill them. A non-zero result is achieved only provided all these
$\delta$-functions are killed.}
\item{Only those terms are taking into account which have the pole $\frac{1}{\varepsilon}$ after
integration over the proper time.}

\end{itemize}

Result of the calculations can be written in the form
\bea
\label{1-div}
\Gamma^{(1)}_{div}[V^{++},Q^{+}] = \frac{C_2-T(R)}{3(4\pi)^3 \epsilon} \tr\int d\zeta^{(-4)} \, (F^{++})^2\\
\nonumber -\frac{2i{\rm f}_{0}^2}{(4\pi)^3} \int d\zeta^{(-4)} \, \tilde{Q}^{+}(C_2 - C(R))Q^{+},
\eea
where $F^{++}=(D^{+})^4V^{--}$, $\epsilon$ is a parameter of dimensional regularization and the
Casimir operators $C_2,\,T(R),\,C(R)$ are defined through the gauge group generators $T^{A}$ as
\bea
\tr(T^{A}T^{B})= T(R)\delta^{AB},\,(T^{A})_{m}{}^{l}(T^{A})_{l}{}^{n}=C(R)\delta_{n}{}^{m},\,
C(Adj)_{m}{}^{n}= C_2\delta_{n}{}^{m}.
\label{Casimir}
\eea
All notations and details of calculations are given in \cite{BIMS-2}. In the case when the hypermultiplet
is in the adjoint representation $T(R)=C(R)=C_2$, the expression (\ref{1-div}) vanishes. Hence,
the $6D,\,\cN=(1,1)$ supergauge theory is off-shell finite at one loop.

\subsection{Two-loop divergences}

Here we discuss a general scheme of singling out the two-loop divergences in $6D,\, \cN=(1,1)$ supergauge theory.
\begin{itemize}
\item{Two-loop divergences are calculated within the background-field
method and the proper-time technique like in one-loop case.}
\item{We deal with the vector multiplet background only.}
\item{The power-counting shows that
the only possible two-loop divergent contribution in the gauge
superfield sector has the following structure \cite{BIMS-4}
$$
\Gamma^{(2)}_{\rm div}[V^{++}] = a \int d\zeta^{(-4)} \, \tr \big( F^{++} \sB
F^{++}\big)\,,
$$
with some constant $a$, which diverges after removing a
regularization.}
\item{Within the background field method, the two-loop contributions to the superfield
effective action are given by the two-loop vacuum harmonic supergraphs with the background-field
dependent lines.}
\item{The background-field dependent propagators (lines) are represented by the proper-time
integrals.}

\end{itemize}

A direct calculation \cite{BIMS-4} yields the following result for
the off-shell divergent part of the two-loop contribution
$\Gamma^{(2)}[V^{++}]$ to effective action in the gauge-superfield
sector
\bea \label{two-loop}
\Gamma^{(2)}_{div} =
\frac{8{\rm f}_{0}^2}{(4\pi)^6\varepsilon^2} (C_2)^2 \mbox{tr} \int
d\zeta^{(-4)}\, F^{++} \sB F^{++}.
\eea
Here we have presented only those
two-loop divergences which contain the two-loop pole
$\frac{1}{\varepsilon^2}$. Calculating the sub-leading divergences
corresponding to the simple pole $\frac{1}{\varepsilon}$ remains an open
issue at present.

The hypermultiplet-dependent contribution to the two-loop divergences can be found by the
straightforward quantum computations of the two-loop effective
action, taking into account the hypermultiplet background $Q^{+}$. However,
the general form of the hypermultiplet-dependent divergences can in
principle be found without direct calculations, just assuming the invariance of the effective action
under the hidden $\cN=(0,1)$ supersymmetry. The result has the extremely simple form
\bea
\Gamma^{(2)}_{\rm div}[V^{++},Q^+]= a\int d \zeta^{(-4)} \, \tr
E^{++}\sB E^{++},
\eea
where $E^{++}=F^{++} + \tfrac{i}2[Q^{+A}, Q^+_A]$ is the classical
equation of motion for the gauge superfield coupled to
the hypermultiplet. It is evident that the two-loop divergences vanish on-shell.
Let us pay attention to the fact that
the two-loop divergences can be canceled by means of the
off-shell redefinition  $V^{++} \rightarrow V^{++} + a \sB
E^{++}$ in the classical action.

\subsection{Low-energy effective action in $6D,\,\cN=(1,1)$ theory}
The structure of finite low-energy
contributions to the one-loop effective action $\Gamma^{(1)}[V^{++}]$ was discussed in refs. \cite{BIM-1,BIM-2}. It was shown that the simplest way to
carry out calculations of the low-energy effective action is to use the formulation of
$6D,\, \cN=(1,1)$ supergauge theory in terms of $\omega$-hypermultiplet \cite{book}. In this case, the
$q$-hypemultiplet-dependent term in the classical action (\ref{S0}) is replaced by
the $\omega$-hypemultiplet term of the form
\bea
\label{omega}
-\frac{1}{2{\rm f}^2} \tr \int d\zeta^{(-4)}\, \nabla^{++}\Omega\nabla^{++}\Omega,
\eea
where ${\rm f}$ is a coupling constant. The first $V^{++}$-dependent term in the action (\ref{S0}) does not change its form. The superfield
$\Omega$ takes values in the adjoint representation of gauge group $SU(N)$ and
$\nabla^{++}\Omega = D^{++}\Omega + i[V^{++},\Omega].$

The effective action $\Gamma[V^{++}]$ is constructed within the background field method, where
the initial superfields $V^{++}$ and $\Omega$ are  split into the background superfields $
\bf{V}^{++},\,\bf{\Omega}$ and the quantum
superfields $v^{++},\, \omega$ by the substitution  $V^{++} \to \bf{V}^{++}$ + {\rm f}$v^{++}$,\,\, $\Omega \to \bf{\Omega}$
+ {\rm f}$\omega$. Details of the background method in this case are given in \cite{BIM-1,BIM-2}.

For further consideration we assume that the background superfields align in a fixed direction inside the
Cartan subalgebra of $su(N)$
\bea
\label{H}
{\bf{V}}^{++} = V_{0}^{++}H,\,\,\,\, {\bf{\Omega}}= \Omega_{0} H,
\eea
where  $H$ is a fixed generator of the Cartan subalgebra which generates some Abelian subgroup $U(1)$.
The choice (\ref{H}) corresponds to the spontaneous symmetry breaking $SU(N) \to SU(N-1)\times U(1)$.
Note that the two background superfields $(V_{0}^{++},\Omega_{0})$ form the Abelian vector $\cN=(1,1)$
multiplet. The bosonic sector of such a multiplet contains a single real gauge vector field $A_M$ and four
real scalars $\phi$ and $\phi^{(ij)}$ with $i,j=1,2$. The fields $A_M,\,\phi,\,\phi^{(ij)}$ in six
dimensions describe the bosonic world-volume degrees of freedom of the $D5$-brane \cite{GK,BKLS}.

The effective action is calculated for the Abelian background superfields $V_{0}^{++},\,\Omega_{0}$ satisfying the
classical equations of motion $F_{0}^{++}=0,\, (D^{++})^2\Omega_{0}=0$ and slowly varying in space-time,
$\partial_{M}W_{0}^{+}=0,\, \partial_{M}\Omega_{0}=0$. Applying the background method and superfield
proper-time technique leads to the leading low-energy correction to the effective action:
\bea
\label{correction}
\Gamma^{(1)}_{lead}= \frac{N-1}{(4\pi)^3}\int d\zeta^{(-4)}\, \frac{(W_{0}^{+})^4}{\Omega_{0}^2}.
\eea
In the bosonic sector the effective action (\ref{correction}) yields
\bea
\Gamma^{(1)}_{bosonic} \sim \int d^{6}x\frac{F^4}{\phi^2}(1 + \frac{\phi^{(ij)}\phi_{(ij)}}{\phi^2}),
\eea
where $F^4 = F_{MN}F^{MN}F_{PQ}F^{PQ} - 4 F^{NM}F_{MR}F^{RS}F_{SN}$ and $F_{MN}$ is the Abelian gauge
strength.

\subsection{One-loop divergences in higher-derivative $6D,\,\cN=(1,0)$ \break supergauge theory}
The higher-derivative $6D,\,\cN=(1,0)$ supersymmetric gauge theory was
formulated in harmonic superspace in ref. \cite{ISZ}.
The classical action of this theory reads
\bea
\label{act-hd} S[V^{++}] = \pm \frac{1}{2g_0^{2}}\tr \int
d\zeta^{(-4)}\, (F^{++})^2,
\eea
where $F^{++}$ was defined in (\ref{identity00}) and $g_{0}$ is a dimensionless coupling constant
\footnote{In the conventional field theory the overall sign of the action is fixed by the requirement of positiveness of energy.
However, in higher-derivative theories the energy is not positively defined in general. Therefore, in the
case under consideration there are no any reasons to fix the overall sign of the action.                                                                  }.
At the component level such a model contains four space-time
derivatives in the bosonic sector. In particular, the component action
includes the term $\sim \int d^6x \tr(\nabla^M F_{MN})^2$, with
$F_{MN}$ being the  standard Yang-Mills strengths. Aspects of the renormaization
of this theory were studied in refs. \cite{ISZ,CasTs}, and
the one-loop counterterms were calculated there in the component approach. The full-fledged
superfield description of the one-loop divergences was given in \cite{BIMS-8,BIMS-9}.

Harmonic superfield quantization of the theory under consideration is carried out in the framework of
the background field method developed for this case in \cite{BIMS-8,BIMS-9}. The method assumes
the background-quantum splitting $V^{++} \to V^{++} + v^{++}$, where $V^{++},\, v^{++}$ in the right-hand
side are background and quantum superfields, respectively, with imposing the gauge-fixing conditions
on $v^{++}$. These conditions are taken to be background-field dependent, in such a way that
the effective action remains gauge invariant. In the one-loop case the effective action is calculated using
the quadratic parts of the classical action and ghost actions. All details can be found in
refs. \cite{BIMS-8,BIMS-9}.

Here we present only the final result for the divergences of the one-loop effective action
\bea
\label{1-loop}
\label{result}
\Gamma^{(1)}_{div} = -\frac{4C_2}{\epsilon (4{\pi}^3} \int d\zeta^{(-4)}\, \tr (F^{++})^2,
\eea
with ${\epsilon}$ being a parameter of dimensional regularization and $C_2$ the second Casimir operator
of $SU(N)$ group. The superfield result (\ref{1-loop}) matches the component calculations
of refs. \cite{ISZ,CasTs}\footnote{The result (\ref{result}) contains contributions from both gauge multiplet and ghost loops (see \cite{BIMS-8,BIMS-9}). }.

Comparing the classical action (\ref{act-hd}) with the one-loop divergent
quantum correction, we can write the one-loop renormalization relation between the bare coupling
constant $g_0$ and the renormalized one $g$ in the form
\bea
\label{renorm}
\frac{1}{g^2} = \frac{1}{g_0^{2}} \mp\frac{22C_2}{3\epsilon(4\pi)^3}.
\eea
The relations (\ref{act-hd}),(\ref{renorm}) imply that the one-loop $\beta$-function in the theory under consideration
has the form
\bea
\label{beta}
\beta(\alpha)= \mp \frac{11\alpha^2 C_2}{24\pi^2},
\eea
where $\alpha = \frac{g^2}{4\pi}$. The lower sign in relation (\ref{beta}) corresponds to Landau pole,
while the upper sign corresponds to asymptotic freedom.

\subsection{Summary and outlook}
Let us summarize. In this Section we presented a brief review of works devoted to studying the quantum structure of six-dimensional
supersymmetric gauge theories. Our study was focused on the development of manifestly supersymmetric and gauge
invariant methods of construction of the effective action in $6D,\, \cN=(1,0)$ and $\cN=(1,1)$ super Yang-Mills
theories and in $6D,\, \cN=(1,0)$ higher-derivative supergauge theory.

The theories under consideration are formulated in six-dimensional harmonic superspace which provides
the manifest $\cN=(1,0)$ supersymmetry. The quantum effective action is defined in the framework of the
background superfield method allowing to carry out the calculations with preserving both $\cN=(1,0)$
supersymmetry and gauge invariance of the effective action. Besides, we developed a power counting
in the harmonic superspace which ensures, along with the manifest $\cN=(1,0)$ supersymmetry and explicit gauge
invariance, the explicit form of possible counterterms up to numerical coefficients.

As a result, we have found the off-shell one-loop divergences and shown that there exists a gauge in which all the
one-loop divergences of $\cN=(1,1)$ theory completely vanish off shell. As the next step, we have studied the
two-loop divergences in this theory. Due to the one-loop finiteness, the two-loop supergraphs do not contain
the divergent subgraphs, which essentially simplifies the analysis of the two-loop counterterms. It has been shown
that at both one-loop and two-loop levels the off-shell counterterms are built only from the classical
equations of motion and, hence, vanish on shell.

Using the harmonic superspace background field method we have developed an approach to construct the low-energy
effective action in $\cN=(1,1)$ supergauge theory. Such an effective action is defined for the
slowly varying background fields corresponding to spontaneous symmetry breaking of the gauge group
$SU(N) \to SU(N-1) \times U(1)$. The calculations were carried out in the framework of the superfield
proper-time technique.

The methods developed  were applied to studying the divergence structure of the one-loop effective action
in a higher-derivative $\cN=(1,0)$ super Yang-Mills theory. In the framework of the off-shell supersymmetric
and gauge invariant approach the one-loop renormalization of the coupling constant was derived and the
corresponding $\beta$-function was found.

The results of the series of our works \cite{BIMS-1,BIMS-2,BIMS-3,BIM-1,BIMS-4,BIMS-5,BIMS-6,BIMS-7,BIM-2,BIMS-8,BIMS-9,BIMS-10,BBIMS,BIMS-11}
clearly demonstrate the power of the harmonic superspace approach for formulation of six-dimensional supersymmetric gauge
theories and for the study of the structure of quantum effective actions of these theories. These
findings suggest a few opportunities to explore a number of open questions related to the structure of effective action in $6D,\,\cN=(1,1)$ supergauge theories.
Firstly, it is important to determine the structure of three-loop divergences, which, unlike one- and two-loop ones, should not disappear on the mass shell. Secondly, it would be extremely interesting to study the possibility of an additional hidden $\cN=(0,1)$
supersymmetry in the gauge multiplet-hypermultiplet higher-derivative system.

\vspace{5mm}\noindent
{\Large\bf Acknowledgements}\nopagebreak
\vspace{0.4cm}

\noindent The authors are grateful to the Academician V.A. Matveev for the suggestion to prepare this review
and to S.A. Fedoruk and D.I. Kazakov for useful comments. We thank B.S. Merzlikin, K.V. Stepanyantz and N.M. Zaigraev for a close collaboration on the topics overviewed here.

\end{document}